# Magnetic properties of ultra-thin 3*d* transition-metal binary alloys I: spin and orbital moments, anisotropy, and confirmation of Slater-Pauling behavior

Martin A. W. Schoen,[1,2]* Juriaan Lucassen,[3] Hans T. Nembach,[1] T. J. Silva,[1] Bert Koopmans[3], Christian H. Back[2] and Justin M. Shaw[1]

[1]Quantum Electromagnetics Division, National Institute of Standards and Technology, Boulder, CO 80305, USA
[2]Institute of Experimental and Applied Physics, University of Regensburg, 93053 Regensburg, Germany
[3]Department of Applied Physics, Eindhoven University of Technology, 5600 MB, Eindhoven, The Netherlands

Dated: 01/05/2017

*Corresponding author: martin1.schoen@physik.uni-regensburg.de

## Abstract

The structure and static magnetic properties - saturation magnetization, perpendicular anisotropy, spectroscopic g-factor, and orbital magnetization - of thin-film 3*d* transition metal alloys are determined over the full range of alloy compositions via X-ray diffraction, magnetometry, and ferromagnetic resonance measurements. We determine the interfacial perpendicular magnetic anisotropy by use of samples sets with varying thickness for specific alloy concentrations. The results agree with prior published data and theoretical predictions. They provide a comprehensive compilation of the magnetic properties of thin-film $Ni_xCo_{1-x}$, $Ni_xFe_{1-x}$ and $Co_xFe_{1-x}$ alloys that goes well beyond the often-cited Slater-Pauling dependence of magnetic moment on alloy concentration.

# 1 Introduction

The magnetic moment of binary 3-d transition metal alloys has been successfully described by the Slater-Pauling model.[1,2] This description is based on a rigid-band model where alloying of a 3$d$ transition metal ferromagnet with another element shifts the Fermi energy, and therefore, the occupation of the magnetic d-states around the Fermi energy. This occupation shift directly translates into a change in magnetization, leading to the basic triangular shape of the Slater-Pauling curve[3–5]. In the $Ni_xFe_{1-x}$ and $Co_xFe_{1-x}$ alloy systems, transitions from a body-centered-cubic (bcc) to a face-centered-cubic (fcc) phase are present. Such phase transitions result in a non-trivial change in the electronic band structure, which can be seen as additional features in the Slater-Pauling curve. For example, a local minimum of magnetic moment typically occurs in the $Ni_xFe_{1-x}$ and $Co_xFe_{1-x}$ alloys at the phase transition.

Despite the fact that the rigid band assumption does not strictly hold for most materials, the Slater-Pauling model[6,7] remains an effective tool for estimating the general behavior of 3-d transition metals. Indeed, further refinement of theoretical descriptions of alloy systems requires that the microscopic band structure is known or calculated, since alloying influences the band structure, for example by smearing out the bands due to atomic disorder[8]. Furthermore, other magnetic properties like orbital magnetization or magnetocrystalline anisotropy cannot be described semi-classically and are purely quantum mechanical in origin.[9]

The development of new magnetic devices such as spin-transfer-torque random-access memory (STT-RAM)[10,11] or spin-torque oscillators (STOs), requires the magnetic properties of each layer to be precisely tuned according to the specifics of the application. The binary alloys of $Ni_xCo_{1-x}$, $Ni_xFe_{1-x}$ and $Co_xFe_{1-x}$ exhibit a wide range of magnetic properties that were thoroughly investigated in bulk samples during the 1960s and 1970s[12–14]. However, many emerging technologies require films of these materials as thin as a few monolayers. Such thin films can exhibit substantially modified magnetic properties from the bulk, including interfacial anisotropy[15,16], strain-induced anisotropy[17–19], reduction of the Curie temperature[20,21], or modification of the magnetic moment[22]. While there have been a number of investigations of materials as thin films,[23–28] there is not yet a comprehensive and systematic study of thin 3-d transition metal alloy films that makes use of modern high-precision characterization methods—such as broad-band ferromagnetic resonance (FMR) or high-resolution X-ray diffraction (XRD). A precise measurement of the magnetic properties for this relatively simple alloy system will facilitate their use in devices, as well as the development of new alloy systems. Here, we report measurements of the alloys $Ni_xCo_{1-x}$, $Ni_xFe_{1-x}$ and $Co_xFe_{1-x}$ over the full range of compositions. We determined the in-plane lattice constant and crystalline structure via XRD, the saturation magnetization density $M_s$ via superconducting quantum interference device (SQUID) magnetometry, as well as the perpendicular anisotropy and the interfacial orbital magnetic moment via FMR. The data presented here are not only a glossary of high-precision measurements of the thin-film magnetic properties in $Ni_xCo_{1-x}$, $Ni_xFe_{1-x}$ and $Co_xFe_{1-x}$, but will also facilitate future testing of predictions based on theoretical calculations, e.g., density functional theory.

# 2 Samples and Method

Thin film samples consisting of $Ni_xCo_{1-x}$, $Ni_xFe_{1-x}$ and $Co_xFe_{1-x}$ alloys were grown at room temperature via dc magnetron sputter deposition on thermally oxidized (001) Si substrates at an Ar pressure of 0.67 Pa ($5\times10^{-3}$ Torr). Substrates were kept in contact with a thermal reservoir, to prevent substrate heating during the deposition process. Film compositions span the full range from

x = 0 to x = 1. The sputter chamber had a base-pressure of less than $5 \times 10^{-6}$ Pa ($4 \times 10^{-8}$ Torr). A Ta(3 nm)/Cu(3 nm) seed layer and Cu(3 nm)/Ta(3 nm) cap layer was used for all samples. The seed layer was chosen to maintain good adhesion to the substrate and promote high quality textured crystalline structure. The capping layer prevents oxidation of the alloy layer, and provides approximately symmetric interfaces and boundary conditions for the excited magnetization. The alloys were co-sputtered from two targets with the deposition rates determined by x-ray reflectivity (XRR). Drift in the deposition rates were periodically monitored with XRR and the repeatability of the deposition rates was found to be better than 3 % over the course of the study. For all deposited alloys, the combined deposition rate was kept at approximately 0.25 nm/s to ensure similar growth conditions. In order to quantitatively account for interfacial effects, we also deposited a thickness series that typically included 10 nm, 7 nm, 4 nm, 3 nm, and 2 nm thicknesses of the pure elements, as well as selected intermediate alloy concentrations ($Ni_{63}Co_{37}$, $Ni_{20}Fe_{80}$, $Ni_{50}Fe_{50}$, $Co_{85}Fe_{15}$, $Co_{50}Fe_{50}$, $Co_{25}Fe_{75}$, and $Co_{20}Fe_{80}$). Following deposition, the samples were coated with ≈150 nm poly(methyl methacrylate) (PMMA) for both mechanical protection and to prevent direct electrical contact to the co-planar waveguide (CPW) used for broadband FMR measurements.

Broadband FMR characterization in the out-of-plane geometry was performed by use of a room temperature bore superconducting magnet capable of applying a perpendicular external magnetic field $H$ as large as $\mu_0 H = 3$ T. Samples were placed face-down on a CPW with a center conductor width of 100 μm with a nominal impedance of 50 Ω. A vector network analyzer (VNA) was connected to both ends of the CPW, and the complex $S_{21}$ transmission parameter (ratio of voltage applied at one end of the CPW to voltage measured at the other end) was measured over a frequency range of 10 GHz to 40 GHz. $S_{21}$ was then fitted with the complex susceptibility tensor component $\chi_{zz}$. For the purpose of fitting, we use

$$S_{21}(H) = A\chi_{zz}(H)e^{i\phi} + mH, \quad (1)$$

with the phase $\phi$ and the dimensionless mode amplitude $A$. A field-dependent complex linear background $mH$ was subtracted to account for measurement drifts. The susceptibility component is derived from the Landau-Lifshitz equation for the perpendicular geometry (z-axis). In the fixed-frequency, field-swept configuration we obtain[29]

$$\chi_{zz}(H) = \frac{M_S(H-M_{\text{eff}})}{(H-M_{\text{eff}})^2-(H_{\text{eff}})^2-i\Delta H(H-M_{\text{eff}})}, \quad (2)$$

where $M_{\text{eff}} = M_S - H_k$ is the effective magnetization, $M_S$ is the saturation magnetization, $H_k$ is the perpendicular anisotropy, and $\Delta H$ is the linewidth. $H_{\text{eff}} = 2\pi f/(\gamma\mu_0)$, where $|\gamma|$ is the gyromagnetic ratio and $\mu_0$ the vacuum permeability. An example of the measured FMR spectra is plotted in Fig. 1, where we present both the real and imaginary parts of $S_{21}$ for $Ni_{90}Fe_{10}$ measured at 20 GHz, in addition to the susceptibility fit to the data.

Both the effective magnetic field $M_{\text{eff}}$ and the spectroscopic g-factor were determined from the resonance field $H_{\text{res}}$ vs. frequency $f$ plot [compare Fig. 1 (c)] according to

$$H_{\text{res}} = M_{\text{eff}} + \frac{h}{g\mu_B\mu_0}f. \quad (3)$$

where $\mu_B$ is the Bohr-magneton. The extracted values for $g$ and $M_{\text{eff}}$ are then corrected for errors stemming from the limited measured frequency range via the method described by Shaw et. al[30].

The crystal structure was characterized by in-plane X-ray diffraction (XRD) using parallel beam optics with a Cu $K_\alpha$ radiation source.

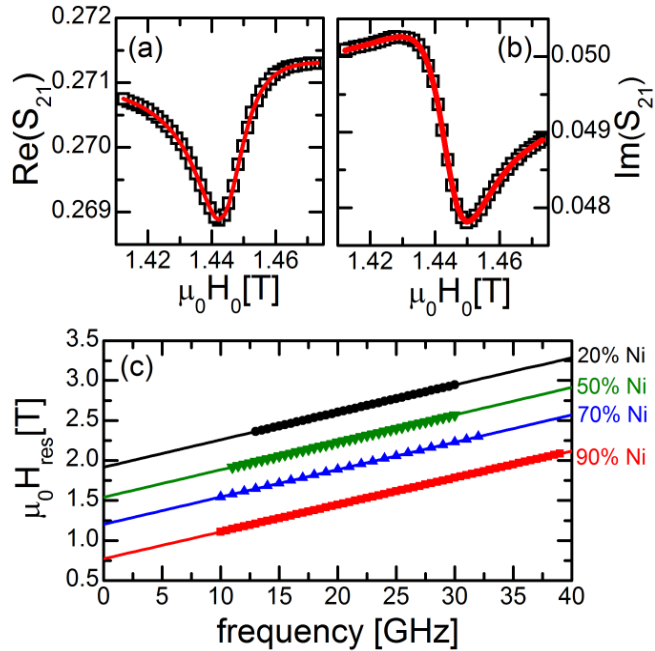

Figure 1: (a) and (b) respectively show the real and imaginary part of the $S_{21}$ transmission parameter (black squares) measured at 20 GHz, along with the complex susceptibility fit (red lines) for the 10 nm $Ni_{90}Fe_{10}$ alloy. In (c), the resonance fields of four $Ni_xFe_{1-x}$ alloys are plotted against the frequency (data points) and fitted linearly (lines) with Ni concentrations denoted on the right axis. The zero-frequency intercepts of the linear fits determine the effective magnetization and their slope is inversely proportional to the g-factor.

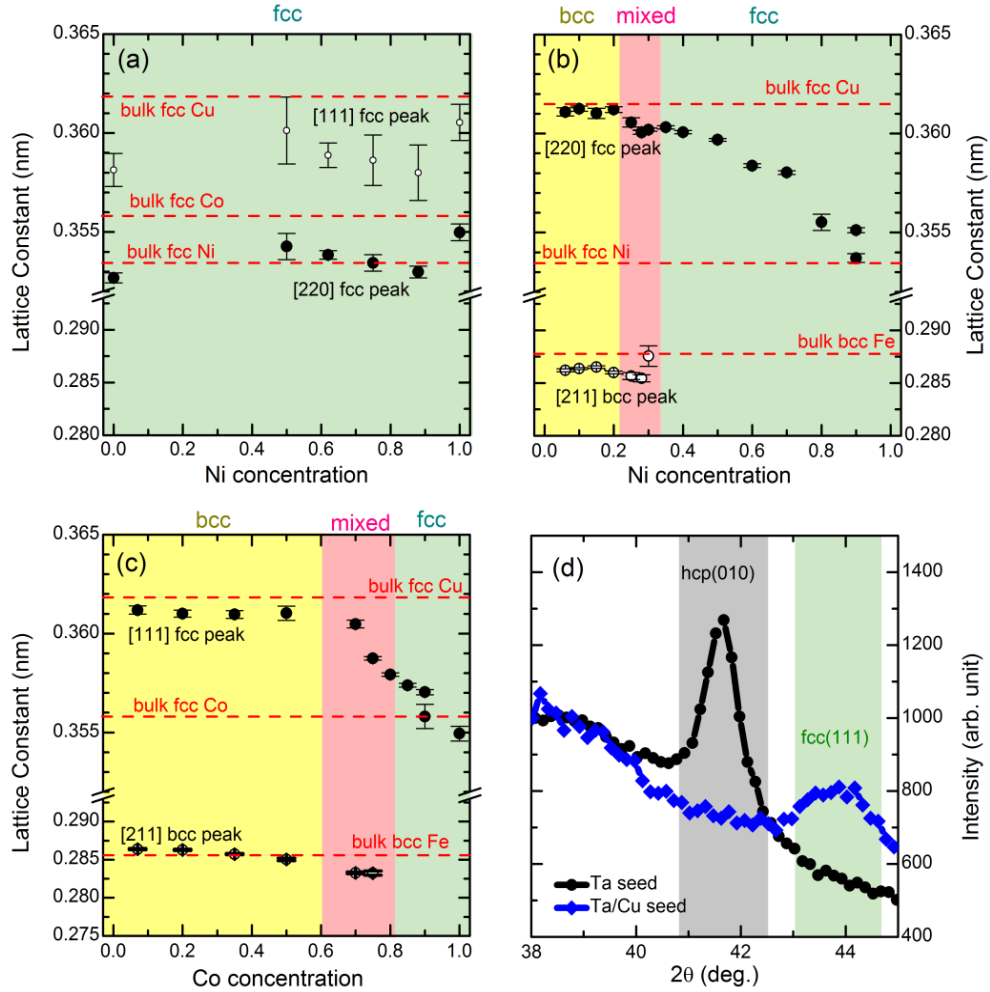

Figure 2: The in-plane lattice constants for (a) $Ni_xCo_{1-x}$, (b) $Ni_xFe_{1-x}$ and (c) $Co_xFe_{1-x}$, as determined by XRD, where the closed circles signify the lattice constant determined from the [220] fcc peak and the open circles either signify the lattice constant from the [111] fcc peak in (a), or from the [211] bcc peak in (b) and (c). Note that the fcc lattice constants are always determined from a superposition of the Cu XRD signal and the fcc alloy XRD signal. The interpretation of multiple peaks is discussed in Sec. 3.1. (d) shows XRD spectra for pure Co grown on both a Ta seed layer and a Ta/Cu seed layer. The Co grown on Ta shows a clear hcp peak, while that grown on Ta/Cu exhibits only a fcc peak.

## 3 Results

### 3.1 XRD

Fig. 2 shows the in-plane lattice constants, determined from the [211] bcc and [111] fcc peaks in the diffraction spectra. The Cu (220) peak is always visible in the spectra due to the Cu seed and cap layers used in all the samples. This complicates the analysis since the location of the fcc-bcc phase transition for the $Ni_xFe_{1-x}$ and $Co_xFe_{1-x}$ alloys cannot be determined exactly. However, the change in the fcc lattice constant away from the bulk Cu value, as well as the diminishment and disappearance of the observed bcc peak, allows us to determine a concentration window in which a mixed phase occurs near the phase transition. Furthermore, we determined the texture of the bcc and fcc phases to be (110) and (111), respectively.

The $Ni_xFe_{1-x}$ alloys exhibit a bcc phase and unchanging bcc lattice constant for a Ni concentration between 0 % and 20 %. For Ni concentrations between 20 % and 30 % a mixed phase[3] is measured as determined by the change in the fcc lattice constant to lower values than for fcc Cu, indicating the formation of a fcc phase in co-existence with a bcc phase, as seen in Fig. 2 (b). The location of this transition is also consistent with the concentration previously reported in the bulk $Ni_xFe_{1-x}$ system at 30 % Ni[3]. For Ni concentrations above 30 % the $Ni_xFe_{1-x}$ alloys exhibit a pure fcc phase, with a lattice constant that approaches the value for pure bulk Ni as the Ni concentration increases.

The $Co_xFe_{1-x}$ system also exhibits a similar fcc-bcc phase transition. The alloys are bcc up to a Co concentration of 60 %, followed by a fcc to bcc phase transition in the vicinity of 70 % Co. This phase transition is again confirmed by XRD, but for the same reason as for the $Ni_xFe_{1-x}$ system, we could only determine that the exact location of the bcc to fcc phase transition occurs somewhere between 60 % and 80 % Co, where bcc and fcc phases co-exist, as shown in Fig. 2 (c). This phase transition seems to occur at a lower Co concentration than for the bulk alloy system[3], which can be attributed to the Cu seed layer, as elaborated in the next paragraph.[33] Above 80 % Co the $Co_xFe_{1-x}$ alloys exhibit purely fcc phase.

Our pure Co films do not exhibit a hexagonal close-packed (hcp) crystal structure, in contrast to prior reports in the literature[3]. Figure Fig. 2 (d) shows the XRD spectrum in the vicinity of the hcp(010) peak for the 10 nm pure Co sample, with a Ta/Cu seed and a Cu/Ta cap layer. For comparison, we include a similar Co film that was grown with only Ta as seed and capping layers. The sample with only Ta in the seed exhibits a clear hcp(010) peak, indicating an hcp structure. In contrast, the sample that includes Cu in the seed and capping layers shows no evidence of hcp structure. We speculate that the strained growth of Co on Cu promotes a strained fcc (i.e., face-centered tetragonal, fct) phase in the Co[31], which is consistent with the structure reported for room-temperature growth of Co/Cu layers via molecular epitaxy[31–33]

The crystalline phase of the $Ni_xCo_{1-x}$ alloys is exclusively fcc and exhibits distinguishable fcc(111) and fcc(220) peaks. The first peak, with its larger lattice constant, can be attributed to Cu, while we attribute the second peak to the $Ni_xCo_{1-x}$.

### 3.2 Magnetization

We determined the room temperature (RT) saturation magnetization $M_S$ for all samples via superconducting quantum interference device (SQUID) magnetometry. The samples were first diced with a precision diamond saw such that the surface area of the sample is accurately known. The saturation magnetization $M_S$ is then determined by dividing the measured magnetic moment by the volume of the magnetic layer. The sample volumes were corrected to account for interfacial factors, like the existence of a dead layer or alloying with the Cu cap and seed layers. Such interfacial effects on the magnetization are determined in a thickness series for select alloys, measured with FMR. The $x$-intercept of the $M_{eff}$ vs. $1/t$ plot (not shown) show a non-zero value of 1.4 nm$^{-1}$, indicating the existence of either a (0.7±0.3) nm magnetic dead layer, or a layer of reduced magnetization corresponding to a thickness of approximately two mono-layers on each interface for all measured samples. The thicknesses of the dead layers display no discernible trend for alloy composition. The existence of a dead layer of this thickness has been found for Fe and $Ni_{80}Fe_{20}$ at similar interfaces[34,35]. Adjusted for the 0.7 nm dead layer, the SQUID measurements were normalized to the alloy volume and the resulting $M_S$ for all alloys is plotted in Fig. 3 (a)-(c) (blue triangles).

For the $Ni_xCo_{1-x}$ alloys, $M_S$ decreases almost linearly with increasing Ni concentration from $\mu_0 M_S(Co) = (1.77 \pm 0.04)$ T to $\mu_0 M_S (Ni) = (0.51 \pm 0.03)$ T. In the $Ni_xFe_{1-x}$ alloy system $M_S$ increases from $\mu_0 M_S (Fe) = (2.05 \pm 0.02)$ T to a maximum of $(2.12 \pm 0.06)$ T at 10 % Ni followed by a minimum at the phase transition (25 % Ni). At Ni concentrations greater than 40 % $M_S$ decreases again with increasing Ni concentration. $M_S$ of the $Co_xFe_{1-x}$ alloys shows a maximum of $(2.42 \pm 0.05)$ T at approximately 35 % Co followed by a decrease with higher Co concentration and a drop at the phase transition.

This behavior for $M_S$ is consistent with the often-observed Slater-Pauling curve[1,2,36], which is included in Fig. 3 as the gray dotted lines. The only deviations occur in the vicinity of the $Ni_xFe_{1-x}$ and $Co_xFe_{1-x}$ phase transitions. In those cases, the "dip" or "drop" in the curve occur at lower Ni or respectively Co concentration than in the bulk Slater-Pauling curve. This is consistent with our XRD measurements that indicate promotion of the fcc phase by the Cu substrate, which causes small deviations relative to the bulk phase diagram.

The effective magnetization $M_{eff}$ is determined by use of Eq. (3) to analyze the FMR data. With the assumption of purely interfacial perpendicular anisotropy, i.e. negligible bulk perpendicular anisotropy, the saturation magnetization $M_S$ can also be determined by measuring $M_{eff}$ in a thickness series (10 nm, 7 nm, 4 nm, 3 nm, 2 nm) and taking the y-intercept (corresponding to infinite thickness $t$) when $M_{eff}$ is plotted versus $1/t$. This is done for a select number of alloys and the resulting values of $M_S$ determined from FMR are also included in Fig. 3 (red crosses). These values for $M_S$ agree well with the values of $M_S$ determined by SQUID, demonstrating the equivalence of both measurement methods.

Furthermore, we determine $M_{eff}$ for the 10 nm thick alloy samples for all concentrations via Eq. 3, with the results plotted in Fig. 3 (black squares) (a)-(c). $M_{eff}$ generally follows the Slater-Pauling curve with an offset due to the presence of interfacial perpendicular anisotropy.

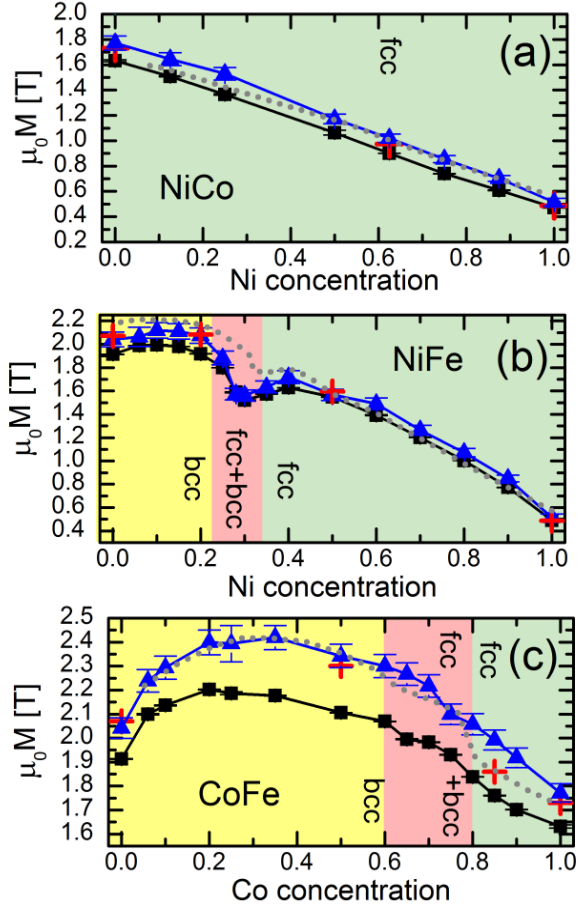

Figure 3: Room temperature effective magnetization $M_{eff}$ (black squares), measured via FMR, and the room temperature saturation magnetization $M_S$ (blue triangles), measured via SQUID magnetometry, are plotted in (a) for $Ni_xCo_{1-x}$, in (b) for $Ni_xFe_{1-x}$ and in (c) for $Co_xFe_{1-x}$. For comparison, $M_S$ is also determined by linear regression of $M_{eff}$ vs. $1/t$ (red crosses). They match $M_S$ by SQUID at those alloy concentrations reasonably well. This shows that the SQUID and FMR measurements are consistent. The crystal structure of the alloys is denoted and signified in the same color code as in Fig. 2. For comparison the bulk "Slater-Pauling" curves for the three alloy systems are also plotted (gray dotted lines)[3].

### 3.3 Perpendicular magnetic anisotropy

As already stated, the difference between $M_{eff}$ from FMR and $M_s$ from magnetometry is the result of interfacial perpendicular anisotropy $H_k$, which results from the broken symmetry at the interfaces[37,38]. Despite being purely interfacial, this interface anisotropy acts on the whole film in the thin film limit and is therefore often employed to engineer the anisotropy according to certain specifications, in particular for perpendicularly magnetized materials[39–42]. Defining the anisotropy energy as in Ref. [43], the effective total perpendicular anisotropy energy density $K$ can be determined from $M_S$ and $M_{eff}$ via[18] $K=1/2(M_S-M_{eff})M_S\mu_0$, and is plotted as a function of alloy concentration in Fig. 4. Note here that we do not separate the different contributions to $K$ (second and fourth order anisotropy constants)[44].

For the $Ni_xCo_{1-x}$ alloy, $K$ decreases almost linearly from pure Co to pure Ni. For the $Ni_xFe_{1-x}$ alloys $K$ has a sharp minimum at the phase transition. We speculate that the minimum is due to the coexistence of multiple phases with compensating amounts of anisotropy. Then, with higher Ni content, $K$ decreases almost linearly.

The $Co_xFe_{1-x}$ alloys behave in a very different manner. Thin films of pure Co and Fe exhibit similar anisotropies, but the alloys have higher values for the anisotropy, up to $2.3 \times 10^5$ J/m$^3$, as shown in Fig. 4 (c). At a Co concentration of 75 % near the fcc to bcc phase transition, $K$ exhibits some degree of distortion.

Under the assumption that $M_S$ is independent of thickness above 2 nm (e.g. Cu does not alloy with the magnetic films), the bulk anisotropy $K_{vol}$ and the average interfacial anisotropy $K_{int}$ can be determined from the thickness dependence of the total anisotropy $K$ by use of the phenomenological equation $K(t)=K_{vol} +2K_{int}/t$ (the factor of 2 accounts for the number of interfaces), [45] where fits of the data based on this equation are presented in the right panels in Fig. 4. We plot the volume and interface components of $K$ with respect to atomic number in Fig. 4 (d) and (e). $K_{vol}$ is small to negligible with no discernable trend with alloy composition. The interface components of the total perpendicular anisotropy for the $Co_{33}Ni_{67}$ alloy is in the range of the one reported by Shaw et. al[45] of $2K_{int}= 1.56\times10^{-4}$ J/m$^2$ for a $(Co_{90}Fe_{10})_{25}Ni_{75}$ alloy with the same seed and cap layers as used in this study.

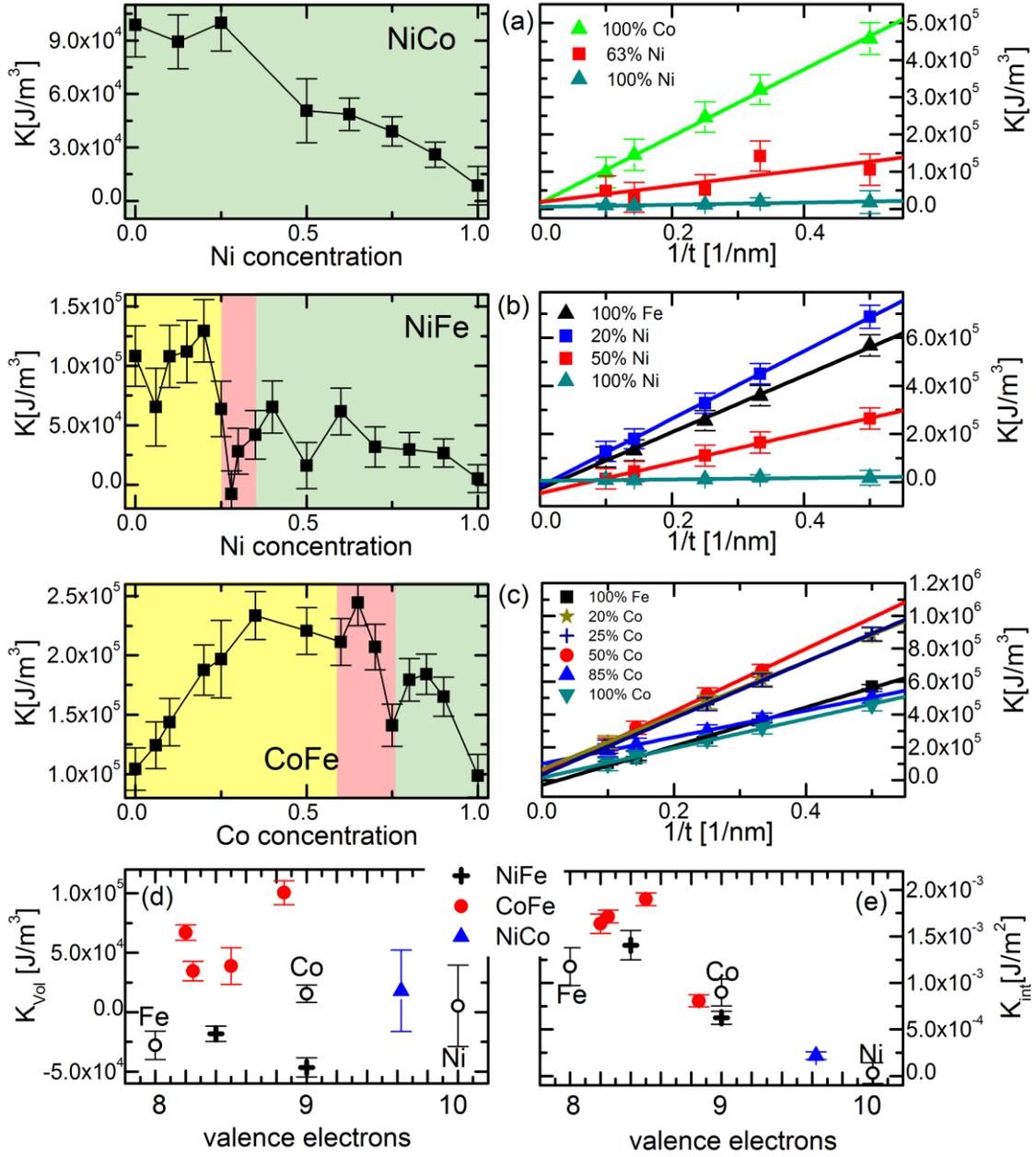

Figure 4: The volume averaged perpendicular anisotropy energy density $K$ is plotted vs. alloy composition for (a) $Ni_xCo_{1-x}$, (b) $Ni_xFe_{1-x}$ and (c) $Co_xFe_{1-x}$. Its thickness dependence for select alloys is plotted in the adjoining panels to the right. Again the crystal structure is signified in color code. The perpendicular anisotropy shows local minimal where fcc and bcc phases coexist for both $Ni_xFe_{1-x}$ and $Co_xFe_{1-x}$. In (d) and (e) we respectively plot the volume averaged bulk anisotropy energy density $K_{vol}$ and the total interfacial anisotropy for both FM/Cu interfaces $K_{int}$, extracted from the intercept and slope via linear regression of $K$ vs. reciprocal thickness $1/t$.

Surprisingly, $K_{int}$ exhibits similar Slater-Pauling behavior as the magnetization data in Fig. 3. This behavior suggests that there is a certain amount of interfacial anisotropy energy per uncompensated d-band spin, i.e., the anisotropy energy is proportional to the spin density at the interface. In Fig. 5 we plot the interface anisotropy against areal spin density and indeed $K_{int}$ increases with the areal spin density. A linear fit to the data yields an $x$ intercept of $(23\pm7)$ $\mu_B/m^2$, which translates to a magnetization of $(0.7\pm0.2)$ $\mu_B$ per interface atom. Considering the symmetry at the 3$d$ transition metal alloy/Cu interface a non-zero $x$ intercept seems reasonable. The symmetry of the localized alloy d-bands is largely broken at the interface with the mostly s-like Cu bands, while the alloy s-p band symmetry should be less affected[46,47]. Thus only the uncompensated localized d-bands should effectively contribute to the perpendicular anisotropy.

While the picture for the itinerant nature of magnetism in the 3$d$ metals remains incomplete, there is substantial evidence that the magnetization in Ni is not found solely in localized d-bands. Tunneling spectroscopy measurements of spin-polarization of the s-p-like conduction electrons for Ni have found values around 23% to 46% (compare Ref. [48] and references therein), which is close to the $x$ intercept value of the linear fit in Fig. 5. It is understood that the s-p bands are strongly hybridized with the d-band near the Fermi surface, giving rise to a high degree of spin polarization for the conduction bands in Ni near the Fermi surface, as revealed by angle resolved photoemission spectroscopy (ARPES)[49]. On the other hand the d-bands in Fe are believed to be more localized[5,50,51]. These considerations side with the value of the $x$ intercept of the linear fit to the data, which is close to $M_s$ of Ni. Furthermore, we can estimate the perpendicular interfacial anisotropy energy per d-band spin to be $(2\pm0.6)\cdot10^{-4}$ eV/$\mu_B$.

Note that the measured interface anisotropies are specifically for the Cu/alloy/Cu interfaces prepared for this study. It is very likely that these anisotropies will also vary with both the choice of non-magnetic metal and the deposition conditions. It is also important to emphasize that the crystalline texture will affect the interfacial anisotropy[52]. In the present case, all of the fcc materials possess a (111) texture, whereas the bcc materials have a (110) texture. But the general trend may indicate a starting point in the search for alloy systems with the desired interface anisotropies.

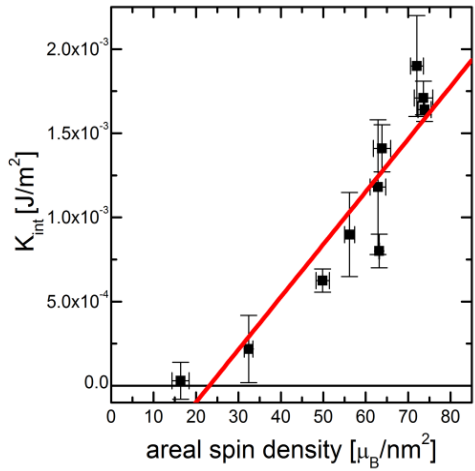

Figure 5: The interfacial anisotropy $K_{int}$ is plotted against the the areal spin density. The red line is a linear fit to the data.

## 3.4 g-factor and orbital magnetization

We now turn to the measured spectroscopic *g*-factor that describes the relationship between the spin angular momentum and total magnetic moment per electron.

For pure Fe and Ni, the *g*-factors are $g_{Fe}$=2.085±0.003 and Ni $g_{Ni}$=2.182±0.016, respectively. For comparison, previously reported values for *bulk* Fe and Ni are $g_{Fe}$=2.08 and $g_{Ni}$=2.185[3,13,53], in good agreement with our results for 10 nm thick films. The agreement between bulk and thin film values is not necessarily expected since there can be a substantial contribution of the orbital moment at the interface[45]. Similarly, the *g*-factor of Co is $g_{Co}$=2.139±0.005, which is very close to the value previously reported for thin film fcc Co $g_{Co}$=2.145[32]. This value is considerably smaller than the one for bulk hcp Co in literature of $g_{Co}$=2.18[3,13]. This is consistent with the XRD results that show no evidence of an hcp phase for the pure Co film. Furthermore, the measured *g*-factor of Permalloy ($Ni_{80}Fe_{20}$) is within 0.2 % of the *g*-factor of $g_{Py}$=2.109 previously reported by Shaw, *et. al* [30]. We found that the g-factor decreases for most alloys with decreasing layer thickness, which has already been observed[30,45]. Interestingly the g-factor increases with decreasing thickness for pure Co and the $Co_{50}Fe_{50}$ alloy.

The *g*-factor for $Ni_xCo_{1-x}$ stays approximately constant for Ni concentrations between 12 % and 66 % after an initial increase from pure fcc Co. At Ni concentrations above 66 %, *g* approaches the value of pure Ni. For comparison, the *g*-factor for hcp Co is also plotted in Fig. 6 (a) and (c). Assuming pure hcp Co a constant *g*-factor of 2.17 is, within a 1 % scatter, a good approximation for the g-factor of all $Ni_xCo_{1-x}$ alloys. The $Ni_xFe_{1-x}$ alloys display a different behavior with Ni-concentration. Starting from pure Fe to $Ni_{80}Fe_{20}$, g only shows an incremental increase, followed by a strong increase in g toward the value for pure Ni. The g-factor in the $Co_xFe_{1-x}$ alloys exhibits a strong non-monotonic behavior. *g* increases with Co concentration from the value for pure Fe and displays a maximum at 10 % Co, followed by a minimum at approximately 20 % Co. With higher Co concentration the alloy *g*-factor increases towards the value for hcp Co and only drops again for pure fcc Co.

We do not observe a strong variation of *g*-factor around the fcc-bcc phase transition of $Ni_xFe_{1-x}$, contrary to the previous report by Bauer and Wigen[12]. Instead, our data for $Ni_xFe_{1-x}$ follow a similar trend as that reported by Meyer and Ash[13]..

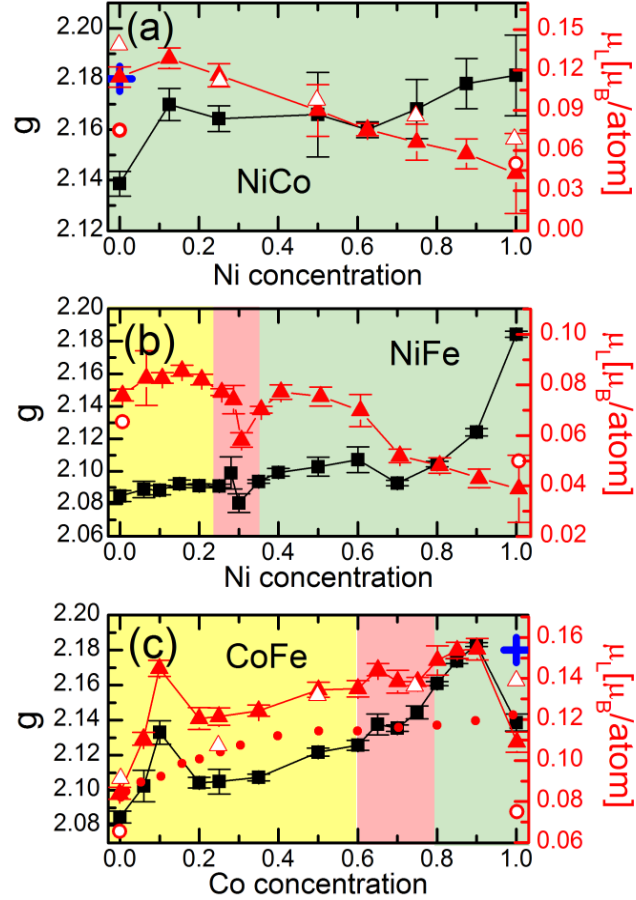

Figure 6: The measured out-of-plane spectroscopic g-factor (black squares, left axis) and the orbital contribution to the saturation magnetization $\mu_L$ (red triangles, right axis) are plotted for (a) $Ni_xCo_{1-x}$, (b) $Ni_xFe_{1-x}$ and (c) $Co_xFe_{1-x}$ against the respective alloy composition (crystal phases are again signified in color). In (a) and (c) the g-factor for pure hcp Co is added[3,13] (blue cross). Orbital moments for the pure elements calculated by Malashevich, et al.[54] are plotted as red open circles for comparison. Also orbital moments for the $Ni_xCo_{1-x}$ and $Co_xFe_{1-x}$ systems calculated by Söderlind, et al.[55] (red open triangles) as well as the orbital moment calculated for $Co_xFe_{1-x}$ by Chadov, et al.[56] (red dots) are included in the respective panels.

The orbital contribution to the magnetization can be calculated from the measured *g*-factor. As a result of the spin-orbit interaction, the g-factor can differ significantly from its undressed value of $\approx 2$. The ratio of orbital $\mu_L$ and spin $\mu_s$ electron moments, to the total magnetic electron moment $\mu$ is given by[57]

$$\frac{\mu_L}{\mu_s} = \frac{g-2}{2}. \quad (4)$$

We use our SQUID magnetometry data to determine the total magnetic moment per atom for each alloy, where we make use of previously published data for alloy atom density[3]. This is used to calculate the spin and orbital moment per atom by use of eq. (4). The atomic orbital moment in Bohr magnetons $\mu_B$ is plotted in Fig. 6 (right axis, red triangles).

Our values agree well with the previous report for the $Ni_xCo_{1-x}$ alloys of Reck and Fry[14]. For $Ni_xFe_{1-x}$, their reported $\mu_L$ is slightly larger than our measured value. It was not possible to discern if the non-monotonic feature exhibited by the $Co_xFe_{1-x}$ alloys were also observed by Reck and Fry due to the density of data in the original report.

Our measurements are consistent with the well-known fact that $\mu_L$ is generally small and on the order of a few percent of the total atomic moment in crystals with cubic symmetry. Beyond that qualitative comparison, the precision of our data enable us to test theoretical *ab initio* models for orbital moments. Theoretical values for $\mu_L$ for pure Ni, Fe and Co[54] reported by Malashevich, *et al.*, are included in Fig. 6 as red open circles. For pure bcc Fe and hcp Co, the calculations yield values that are approximately 30 % lower than our experimentally determined values, whereas for fcc Ni the theory values are larger than the experimental values of $\mu_L$. The results of Söderlind, *et al.*[55] for the $Ni_xCo_{1-x}$ and $Co_xFe_{1-x}$ alloys are included as red open triangles, and the theoretical predictions of Chadov, *et al.*[56] for the $Co_xFe_{1-x}$ system are included as red dots. The predictions of Söderlind, *et al.* for the $Ni_xCo_{1-x}$ and $Co_xFe_{1-x}$ alloys are remarkably close to our measured values, with the possible exception of pure Co, where we measured $\mu_L = 0.11$ $\mu_B$, and they predicted 0.14 $\mu_B$. The calculated results of Chadov, *et al.*, match our measured values for pure Co and Fe quite well, but they are significantly lower than the measured values for all the alloys. The model also fails to capture the substantial jump in the orbital moment with the addition of Co at low concentrations (< 10 %), and the similarly precipitous drop as the alloy approaches pure Co.

## 4 Summary

We present a comprehensive study on the crystalline structure, effective magnetization, saturation magnetization, perpendicular anisotropy, *g*-factor and orbital magnetic moment for 10 nm thick binary alloys $Ni_xCo_{1-x}$, $Ni_xFe_{1-x}$ and $Co_xFe_{1-x}$ over the full range of alloy compositions. The measured saturation magnetization is consistent with the Slater-Pauling behavior for bulk specimens. By measuring the effective magnetization via FMR and the saturation magnetization via SQUID magnetometry, we calculate the perpendicular magnetic anisotropy energy density for all compositions. For a subset of alloy concentrations, we determine the bulk and interfacial contributions to the perpendicular magnetic anisotropy. While the bulk anisotropy energy density shows no discernable trend with alloy concentration, the interfacial contribution also exhibits Slater-Pauling-like behavior, which implies a fixed amount of interfacial anisotropy energy per localized, uncompensated, *d*-band spin. The measured *g*-factor agrees well with previously published results for the pure elements. Furthermore, we determine the orbital magnetic moments for all the alloys. Many of our measured values for $\mu_L$ are in good agreement with previous predictions that were obtained from *ab initio* calculations.

# 5 References

1. Pauling, L. The Nature of the Interatomic Forces in Metals. *Phys. Rev.* **54**, 899–904 (1938).
2. Slater, J. C. Electronic Structure of Alloys. *Journal of Applied Physics* **8**, 385–390 (1937).
3. Bozorth, R. M. *Ferromagnetism*. (IEEE Press: 2003).
4. Kakehashi, Y. *Modern Theory of Magnetism in Metals and Alloys*. (Springer-Verlag Berlin Heidelberg: 2012).
5. Kübler, J. *Theory of Itinerant Electron Magnetism*. (Oxford University Press Inc., New York: 2000).
6. Malozemoff, A. P., Williams, A. R. & Moruzzi, V. L. Band-gap theory" of strong ferromagnetism: Application to concentrated crystalline and amorphous Fe- and Co-metalloid alloys. *Phys. Rev. B* **29**, 1620–1632 (1984).
7. Williams, A., Moruzzi, V., Malozemoff, A. & Terakura, K. Generalized Slater-Pauling curve for transition-metal magnets. *IEEE Transactions on Magnetics* **19**, 1983–1988 (1983).
8. Schoen, M. A. W. *et al.* Ultra-low magnetic damping of a metallic ferromagnet. *Nat. Phys* **12**, 839–842 (2016).
9. *Ultrathin Magnetic Structures I*. (Springer-Verlag Berlin Heidelberg: 1994).
10. Amiri, P. K. *et al.* Low Write-Energy Magnetic Tunnel Junctions for High-Speed Spin-Transfer-Torque MRAM. *Electron Device Letters, IEEE* **32**, 57–59 (2011).
11. Kaka, S. *et al.* Spin transfer switching of spin valve nanopillars using nanosecond pulsed currents. *Journal of Magnetism and Magnetic Materials* **286**, 375–380 (2005).
12. Bauer, C. A. & Wigen, P. E. Spin-Wave Resonance Studies in Invar Films. *Phys. Rev. B* **5**, 4516–4524 (1972).
13. Meyer, A. J. P. & Asch, G. Experimental g' and g Values of Fe, Co, Ni, and Their Alloys. *Journal of Applied Physics* **32**, S330–S333 (1961).
14. Reck, R. A. & Fry, D. L. Orbital and Spin Magnetization in Fe-Co, Fe-Ni, and Ni-Co. *Phys. Rev.* **184**, 492–495 (1969).
15. El Gabaly, F. *et al.* Imaging Spin-Reorientation Transitions in Consecutive Atomic Co Layers on Ru(0001). *Phys. Rev. Lett.* **96**, 147202 (2006).
16. Farle, M., Platow, W., Anisimov, A. N., Poulopoulos, P. & Baberschke, K. Anomalous reorientation phase transition of the magnetization in fct Ni/Cu(001). *Phys. Rev. B* **56**, 5100–5103 (1997).
17. Prokop, J. *et al.* Strain-induced magnetic anisotropies in Co films on Mo(110). *Phys. Rev. B* **70**, 184423 (2004).
18. Rizal, C., Gyawali, P., Kshattry, I. & Pokharel, R. K. Strain-induced magnetoresistance and magnetic anisotropy properties of Co/Cu multilayers. *Journal of Applied Physics* **111**, 07C107 (2012).
19. Sander, D. The correlation between mechanical stress and magnetic anisotropy in ultrathin films. *Reports on Progress in Physics* **62**, 809 (1999).
20. Schneider, C. M. *et al.* Curie temperature of ultrathin films of fcc-cobalt epitaxially grown on atomically flat Cu(100) surfaces. *Phys. Rev. Lett.* **64**, 1059–1062 (1990).
21. Sun, L., Searson, P. C. & Chien, C. L. Finite-size effects in nickel nanowire arrays. *Phys. Rev. B* **61**, R6463–R6466 (2000).
22. García-Arribas, A. *et al.* Tailoring the magnetic anisotropy of thin film permalloy microstrips by combined shape and induced anisotropies. *The European Physical Journal B* **86**, 1–7 (2013).
23. Bayreuther, G., Dumm, M., Uhl, B., Meier, R. & Kipferl, W. Magnetocrystalline volume and interface anisotropies in epitaxial films: Universal relation and Néel's model (invited). *Journal of Applied Physics* **93**, 8230–8235 (2003).
24. Bayreuther, G. & Lugert, G. Magnetization of ultra-thin epitaxial Fe films. *Journal of Magnetism and Magnetic Materials* **35**, 50–52 (1983).
25. Dumm, M. *et al.* Magnetism of ultrathin FeCo (001) films on GaAs(001). *Journal of Applied Physics* **87**, 5457–5459 (2000).
26. Heinrich, B. *et al.* Development of magnetic anisotropies in ultrathin epitaxial films of Fe(001) and Ni(001). *Applied Physics A* **49**, 473–490 (1989).
27. Huang, F., Kief, M. T., Mankey, G. J. & Willis, R. F. Magnetism in the few-monolayers limit: A surface magneto-optic Kerr-effect study of the magnetic behavior of ultrathin films of Co, Ni, and Co-Ni alloys on Cu(100) and Cu(111). *Phys. Rev. B* **49**, 3962–3971 (1994).
28. Reiger, E. *et al.* Magnetic moments and anisotropies in ultrathin epitaxial Fe films on ZnSe(001). *Journal of Applied Physics* **87**, 5923–5925 (2000).
29. Nembach, H. T. *et al.* Perpendicular ferromagnetic resonance measurements of damping and Lande g-factor in sputtered (Co$_2$Mn$_{1-x}$Ge$_x$ films. *Phys. Rev. B* **84**, 054424 (2011).

30. Shaw, J. M., Nembach, H. T., Silva, T. J. & Boone, C. T. Precise determination of the spectroscopic g-factor by use of broadband ferromagnetic resonance spectroscopy. *Journal of Applied Physics* **114**, 243906 (2013).
31. Kief, M. T. & Egelhoff, W. F. Growth and structure of Fe and Co thin films on Cu(111), Cu(100), and Cu(110): A comprehensive study of metastable film growth. *Phys. Rev. B* **47**, 10785–10814 (1993).
32. Pelzl, J. *et al.* Spin-orbit-coupling effects on g-value and damping factor of the ferromagnetic resonance in Co and Fe films. *Journal of Physics: Condensed Matter* **15**, S451 (2003).
33. Tischer, M. *et al.* Enhancement of Orbital Magnetism at Surfaces: Co on Cu(100). *Phys. Rev. Lett.* **75**, 1602–1605 (1995).
34. Lambert, C.-H. *et al.* Quantifying perpendicular magnetic anisotropy at the Fe-MgO(001) interface. *Applied Physics Letters* **102**, 122410 (2013).
35. Marko, D., Strache, T., Lenz, K., Fassbender, J. & Kaltofen, R. Determination of the saturation magnetization of ion irradiated Py/Ta samples using polar magneto-optical Kerr effect and ferromagnetic resonance. *Applied Physics Letters* **96**, 022503 (2010).
36. Morrish, A. H. *The Physical Principles of Magnetism*. (IEEE Press: 2001).
37. Bruno, P. Magnetic surface anisotropy of cobalt and surface roughness effects within Neel's model. *J. Phys. F: Met. Phys.* **18**, 1291–1298 (1988).
38. Neel, L. L'approche á la saturation de la magnétostriction. *J. Phys. Radium* **15**, 376–378 (1954).
39. Bruno, P. & Renard, J.-P. Magnetic surface anisotropy of transition metal ultra thin films. *Appl. Phys. A* **49**, 499–506 (1989).
40. Johnson, M. T., Bloemen, P. J. H., Broeder, F. J. A. den & Vries, J. J. de Magnetic anisotropy in metallic multilayers. *Rep. Prog. Phys.* **59**, 1409–1458 (1996).
41. Shaw, J. M., Nembach, H. T. & Silva, T. J. Damping phenomena in Co_90Fe_10/Ni multilayers and alloys. *Applied Physics Letters* **99**, (2011).
42. Watanabe, M., Homma, M. & Masumoto, T. International Conference on Magnetism (Part II) Perpendicularly magnetized Fe-Pt (0 0 1) thin films with (B • H)$_{max}$ exceeding 30 MG Oe. *Journal of Magnetism and Magnetic Materials* **177**, 1231–1232 (1998).
43. Farle, M., Mirwald-Schulz, B., Anisimov, A. N., Platow, W. & Baberschke, K. Higher-order magnetic anisotropies and the nature of the spin-reorientation transition in face-centered-tetragonal Ni(001)/Cu(001). *Phys. Rev. B* **55**, 3708–3715 (1997).
44. *Ultrathin Magnetic Structures II*. (Springer-Verlag Berlin Heidelberg: 1994).
45. Shaw, J. M., Nembach, H. T. & Silva, T. J. Measurement of orbital asymmetry and strain in Co_90Fe_10/Ni multilayers and alloys: Origins of perpendicular anisotropy. *Phys. Rev. B* **87**, 054416 (2013).
46. Újfalussy, B., Szunyogh, L. & Weinberger, P. Magnetic anisotropy in Fe/Cu(001) overlayers and interlayers: The high-moment ferromagnetic phase. *Phys. Rev. B* **54**, 9883–9890 (1996).
47. Nilsson, A. *et al.* Determination of the electronic density of states near buried interfaces: Application to Co/Cu multilayers. *Phys. Rev. B* **54**, 2917–2921 (1996).
48. Meservey, R. & Tedrow, P. M. Spin-polarized electron tunneling. *Physics Reports* **238**, 173–243 (1994).
49. Altmann, K. N. *et al.* Enhanced spin polarization of conduction electrons in Ni explained by comparison with Cu. *Phys. Rev. B* **61**, 15661–15666 (2000).
50. Carbone, C. *et al.* Correlated Electrons Step by Step: Itinerant-to-Localized Transition of Fe Impurities in Free-Electron Metal Hosts. *Phys. Rev. Lett.* **104**, 117601 (2010).
51. Knut, R. *et al.* Localization of Fe d-states in Ni-Fe-Cu alloys and implications for ultrafast demagnetization. *arXiv:1508.03015* (2015).
52. Engel, B. N., England, C. D., Van Leeuwen, R. A., Wiedmann, M. H. & Falco, C. M. Interface magnetic anisotropy in epitaxial superlattices. *Phys. Rev. Lett.* **67**, 1910–1913 (1991).
53. Maeda, T., Yamauchi, H. & Watanabe, H. Spin Wave Resonance and Exchange Parameters in fcc Fe-Ni Alloys. *Journal of the Physical Society of Japan* **35**, 1635–1642 (1973).
54. Malashevich, A., Souza, I., Coh, S. & Vanderbilt, D. Theory of orbital magnetoelectric response. *New Journal of Physics* **12**, 053032 (2010).
55. Söderlind, P., Eriksson, O., Johansson, B., Albers, R. C. & Boring, A. M. Spin and orbital magnetism in Fe-Co and Co-Ni alloys. *Phys. Rev. B* **45**, 12911–12916 (1992).
56. Chadov, S. *et al.* Orbital magnetism in transition metal systems: The role of local correlation effects. *EPL* **82**, 37001 (2008).
57. Kittel, C. On the Gyromagnetic Ratio and Spectroscopic Splitting Factor of Ferromagnetic Substances. *Phys. Rev.* **76**, 743–748 (1949).